\documentclass[showpacs,pra,twocolumn,amsmath]{revtex4}
\usepackage{graphicx}

\begin{document}
\title{Local entanglement and quantum phase transition in 1D transverse field Ising model}

%\footnote[1]{A brief summary of results in this paper has been present at a
%recent ``International Conference on the Frontiers of Nonlinear and Complex
%Systems", 24-26 May 2006, Hong Kong, http://www.hkbu.edu.hk/\~~cns10.}

\author{Shi-Quan Su}
\author{Jun-Liang Song}
\author{Shi-Jian Gu}
\altaffiliation{Email: sjgu@phy.cuhk.edu.hk\\URL: http://www.phystar.net/}
\affiliation{Department of Physics and Institute of Theoretical Physics, The
Chinese University of Hong Kong, Hong Kong, China}

\begin{abstract}
In this paper, we study the entanglement between two-neighboring sites and the
rest of the system in a simple quantum phase transition of 1D transverse field
Ising model. We find that the entanglement shows interesting scaling and
singular behavior around the critical point, and then can be use as a convenient
marker for the transition point.

\end{abstract}
\date{\today}
\maketitle

Entanglement is a one of the most fascinating features \cite{ABinstein35} of the
quantum theory. It roots in the superposition principle of quantum mechanics and
the tensor product structure of different Hilbert spaces, and implies the
existence nonlocal correlation in quantum world. Since this correlation is
absent in the classical world, it is a kind of pure quantum correlation and
regarded as crucial resource in many quantum information processing tasks
\cite{SeeForExample,MANielsenb}. Actually, similar fascinating property also
exists in the classic optics. By recording the interference pattern of two laser
beams reflected from one object, the whole image can be recovered from even a
small piece of holograph, although the resolution may be reduced. In other
words, the superposition of two coherent electromagnetic waves enables us to
learn some global information from a localized spatial area. Though it is not a
precise analogy, people expect that the entanglement, which roots in the same
superposition principle, can enable us to learn some global properties from a
small part of the system.  This observation may be one of the main motivations
in the recent studies
\cite{SJGUPRL,anfossi2005,larsson2005,anfossi2006,legeza2006,gu2006le,
venuti2006,GVidal2003,VEKorepin04,JCao06} on the role of entanglement between a
small part, e.g. a block consisting of one or more sites, and the rest of the
system in the quantum phase transition \cite{Sachdev}.

Conceptually, the single-site and the two-site entanglement, and the
block-block entanglement \cite{GVidal2003,VEKorepin04,YChenNJP} are different
notations of the same kind of measurement. They are defined as the von Neumann
entropy of the reduced density matrix of a local part in the system. Studies of
the block-block entanglement in the one-dimensional spin models have
established a close relationship between the conformal field theory and quantum
information theory \cite{VEKorepin04}. The block-block entanglement is also
generalized in systems other than spin models. Many works are devoted to
understand the local entanglement, a limiting case of the block-block
entanglement, in the ground state of itinerant fermion systems. For example, in
the extended Hubbard model, the global phase diagram can be sketched out by the
contour map of the single-site local entanglement in the parameter space
\cite{SJGUPRL}. And the entanglement's scaling behavior is investigated in the
one-dimensional Hubbard model at criticality point \cite{larsson2005}. In the
Hirsch model, the singularity in the first derivative of the single-site
entanglement is used to locate the superconductor-insulator transition point
\cite{anfossi2006}. However, in some spin systems with $Z2$ symmetry, the
single-site entanglement usually does not contain enough information to
describe the critical behavior \cite{legeza2006,gu2006le}, then it is necessary
to include more sites, such as two sites in the simplest case, into the block.
For example, in 1D XXZ model, the two-site local entanglement reaches maximum
at the isotropic transition point \cite{gu2006le}.

These results suggest that the local entanglement may be used as a good and
convenient marker of quantum phase transition. In a strongly correlated system,
the reduced density matrix of a local block involves the bi-partite correlation
in the system as the consequence of the superposition principle in quantum
mechanics. At the quantum critical point, quantum fluctuations extend to all
scales, thus it may contribute a singular part in the local entanglement. The
idea of using the local entanglement to locate transition point is especially
useful in numerical calculations \cite{anfossi2006}, usually the data obtained
from a relative small size lattice is sufficient to detect the quantum phase
transitions.

In this paper, we investigate the scaling behavior of the two-site local
entanglement near the quantum phase transition point in 1D transverse field
Ising model. Since the transition type of the Ising model is different from that
in 1D XXZ model, the entanglement shows quite different behavior, such as
scaling and singularity. We calculate the entanglement and its first order
derivative with respect to the coupling constant numerically. Based on the exact
solution \cite{EBarouch70,EBarouch71}, we also obtain explicit expressions which
clearly exhibit the logarithmic divergence of local entanglement's derivative at
the critical point. Therefore, by studying the two-site local entanglement, we
hope to have a deep understanding on the another type of quantum phase
transition, as represented by 1D Ising model.

The Hamiltonian of 1D transverse field Ising model reads
\begin{eqnarray}
&&H_{\rm Ising
}=-\sum_{j=1}^N\left[\lambda\sigma_j^x\,\sigma_{j+1}^x+\sigma_j^z\right],
\nonumber
\\ && \sigma_1=\sigma_{N+1}, \label{eq:hamitl2}
\end{eqnarray}
where $\sigma_i (\sigma^x, \sigma^y, \sigma^z)$ are Pauli matrices at site $i$,
$\lambda$ is an Ising coupling in unit of the transverse field, and the periodic
boundary conditions are assumed. In the basis spanned by the eigenstates of
$\{\sigma^z\}$, the Hamiltonian changes the number of down spins by two due to
the term $\sigma^x_i\sigma^x_{i+1}$ in its expression, so the whole space of
system can be divided by the parity of the number of down spins. That is the
Hamiltonian and the parity operator $P=\prod_j\sigma_j^z$ can be simultaneously
diagonalized and the eigenvalues of $P$ is $\pm 1$. Then it can be proved that
for a finite system, the ground state in the whole region of $\lambda>0$ is
non-degenerate.

We confine our interest to the entanglement between two neighboring sites and
rest of the system, so we need to consider the reduced density matrix of two
local sites. According to the parity conservation, the reduced density matrix of
two spins in this system takes the form
\begin{equation}
{\rho}_{ij} = \left(
\begin{array}{llll}
u^+ & 0 & 0 & z^- \\
0 & w_1 & z^+ & 0 \\
0 & z^+ & w_2 & 0 \\
z^- & 0 & 0 & u^-
\end{array}
\right)
\end{equation}
in the basis $\{|\uparrow\uparrow\rangle, |\uparrow\downarrow\rangle,
|\downarrow\uparrow\rangle, |\downarrow\downarrow\rangle\}$. The elements in the
density matrix $\rho_{ij}$ can be calculated from the correlation functions
\begin{eqnarray}
&&u^\pm=\frac{1}{4}(1\pm2\langle\sigma_i^z \rangle
+\langle\sigma_i^z\sigma_j^z\rangle), \nonumber \\
&&w_1=w_2=\frac{1}{4}(1- \langle\sigma_i^z\sigma_j^z\rangle), \nonumber \\
&&z^\pm = \frac{1}{4}(\langle\sigma_i^x\sigma_j^x\rangle
\pm\langle\sigma_i^y\sigma_j^y\rangle). \label{eq:eleinRDM}
\end{eqnarray}
In a finite system, the ground state is a pure state, so the von Neumann entropy
$E_v$, i.e.
\begin{eqnarray}
E_v=-{\rm tr}\left[\rho_{ij}\log_2(\rho_{ij})\right]
\end{eqnarray}
measures the entanglement between the sites $i,j$ and the rest $N-2$ sites of
the system.

\begin{figure}
\includegraphics[width=8cm]{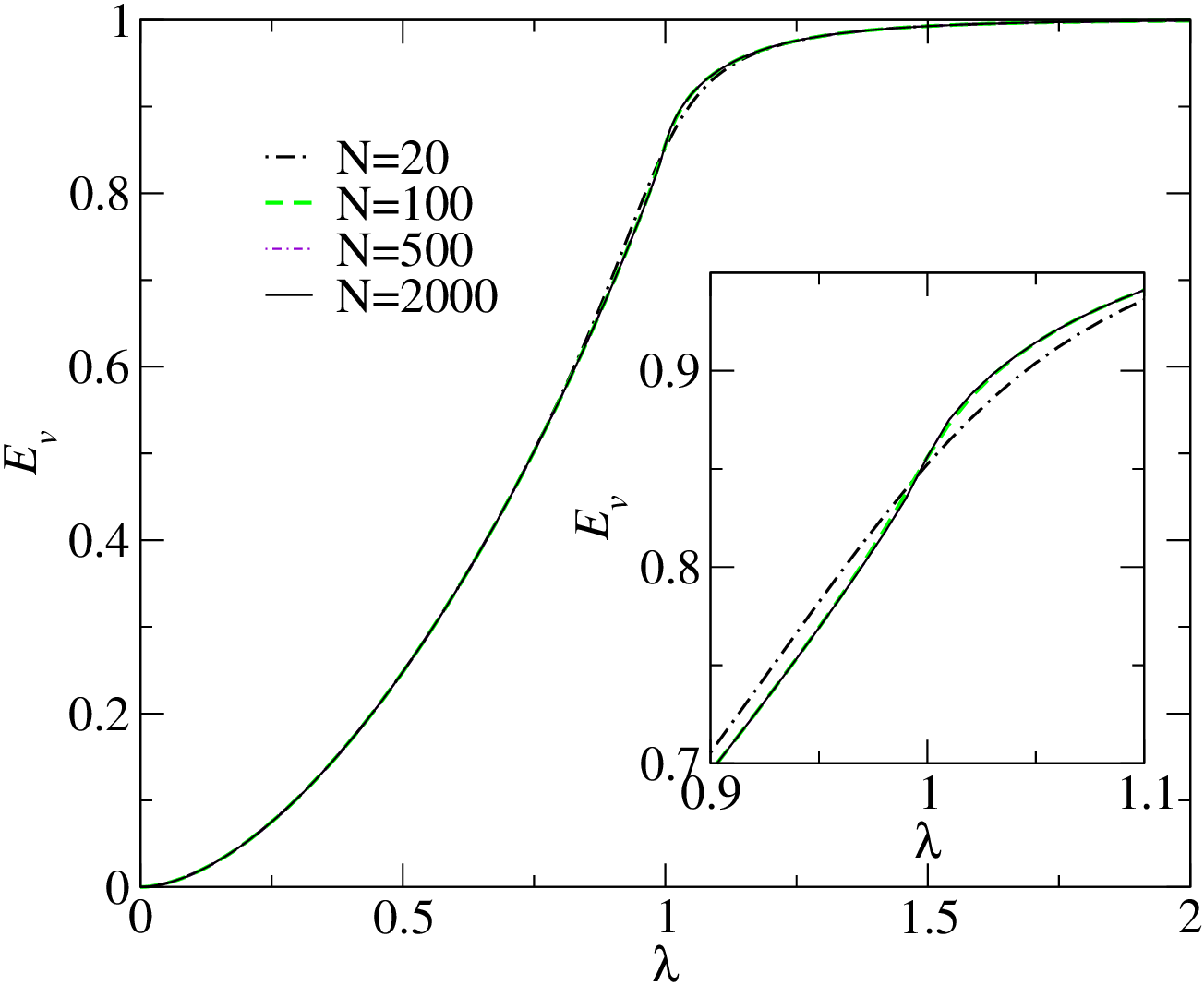}
\caption{\label{figure_lent}(color online) The two-site local entanglement as a
function of $\lambda$ for various system size.}
\end{figure}

\begin{figure}
\includegraphics[width=8cm]{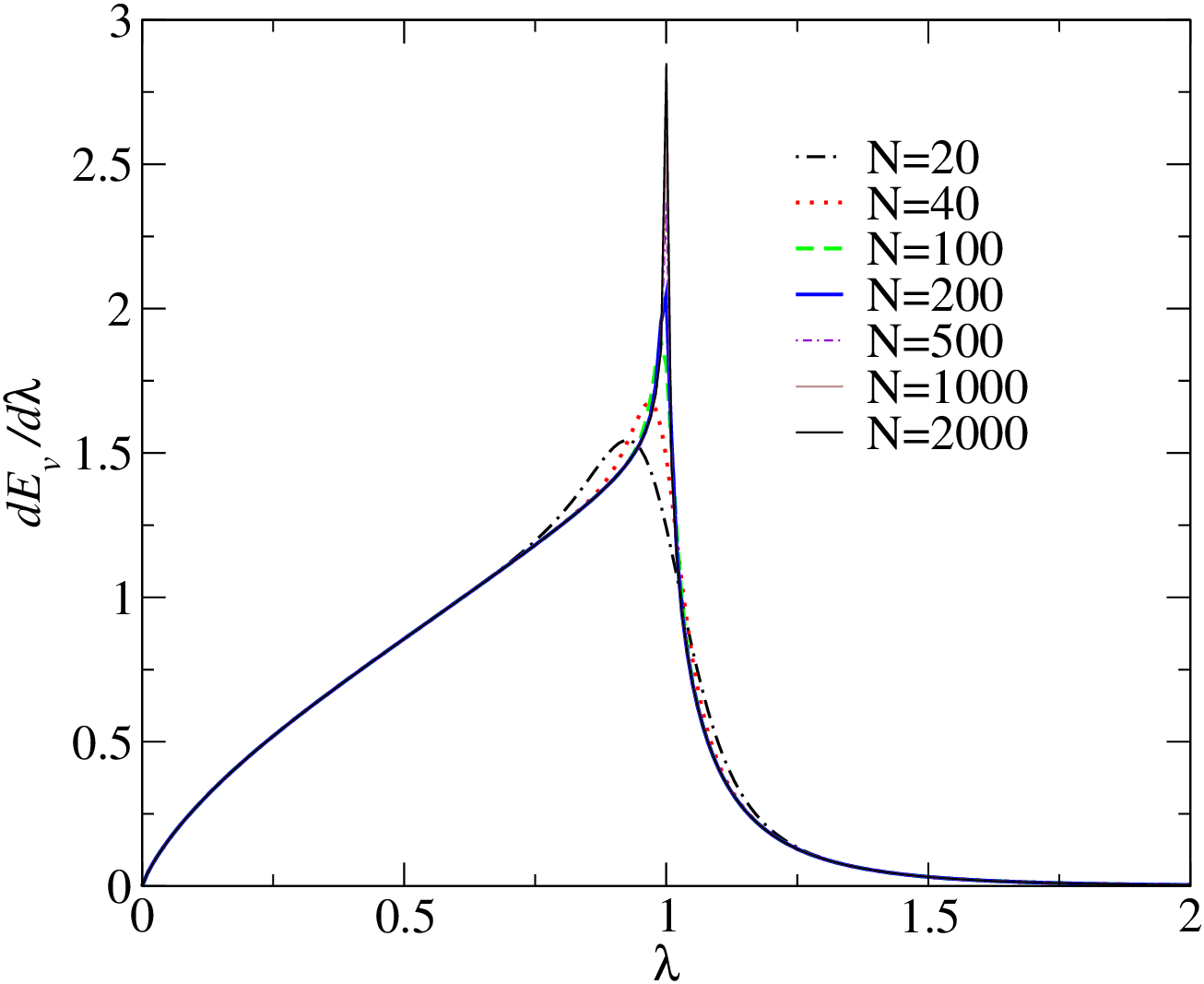}
\caption{\label{figure_dlent}(color online) The first order derivative of the
two-site local entanglement as a function of $\lambda$ for various system size.}
\end{figure}

The correlation functions in Eqs (\ref{eq:eleinRDM}) can be calculated exactly.
The mean magnetization in the ground state is given by \cite{EBarouch70}
\begin{eqnarray}
\langle\sigma^z\rangle = \frac{1}{N}\sum_\phi
\frac{(1-\lambda\cos\phi)}{\omega_\phi},
\end{eqnarray}
where $\omega_\phi$ is the dispersion relation,
\begin{eqnarray}
&&\omega_\phi=\sqrt{1+\lambda^2-2\lambda\cos(\phi_q)},\nonumber \\ &&\phi_q=2\pi
q/N,
\end{eqnarray}
where $q$ is integer (half-odd integer) for parity $P=-1 (+1)$. The two-point
correlation functions are calculated as \cite{EBarouch71}
\begin{eqnarray}
\langle\sigma_0^x\sigma_1^x\rangle &=& \frac{1}{N}\sum_\phi
\frac{\lambda-\cos\phi
}{\omega_\phi}\nonumber \\
\langle\sigma_0^y\sigma_1^y\rangle &=& \frac{1}{N}\sum_\phi
\frac{\lambda\cos(2\phi)-\cos\phi }{\omega_\phi} \nonumber \\
\langle\sigma_0^z\sigma_1^z\rangle
&=&\langle\sigma^z\rangle^2-\langle\sigma_0^x\sigma_1^x\rangle
\langle\sigma_0^y\sigma_1^y\rangle
\end{eqnarray}
Therefore, we can calculate the two-site local entanglement for arbitrary finite
system directly.

\begin{figure}
\includegraphics[width=8cm]{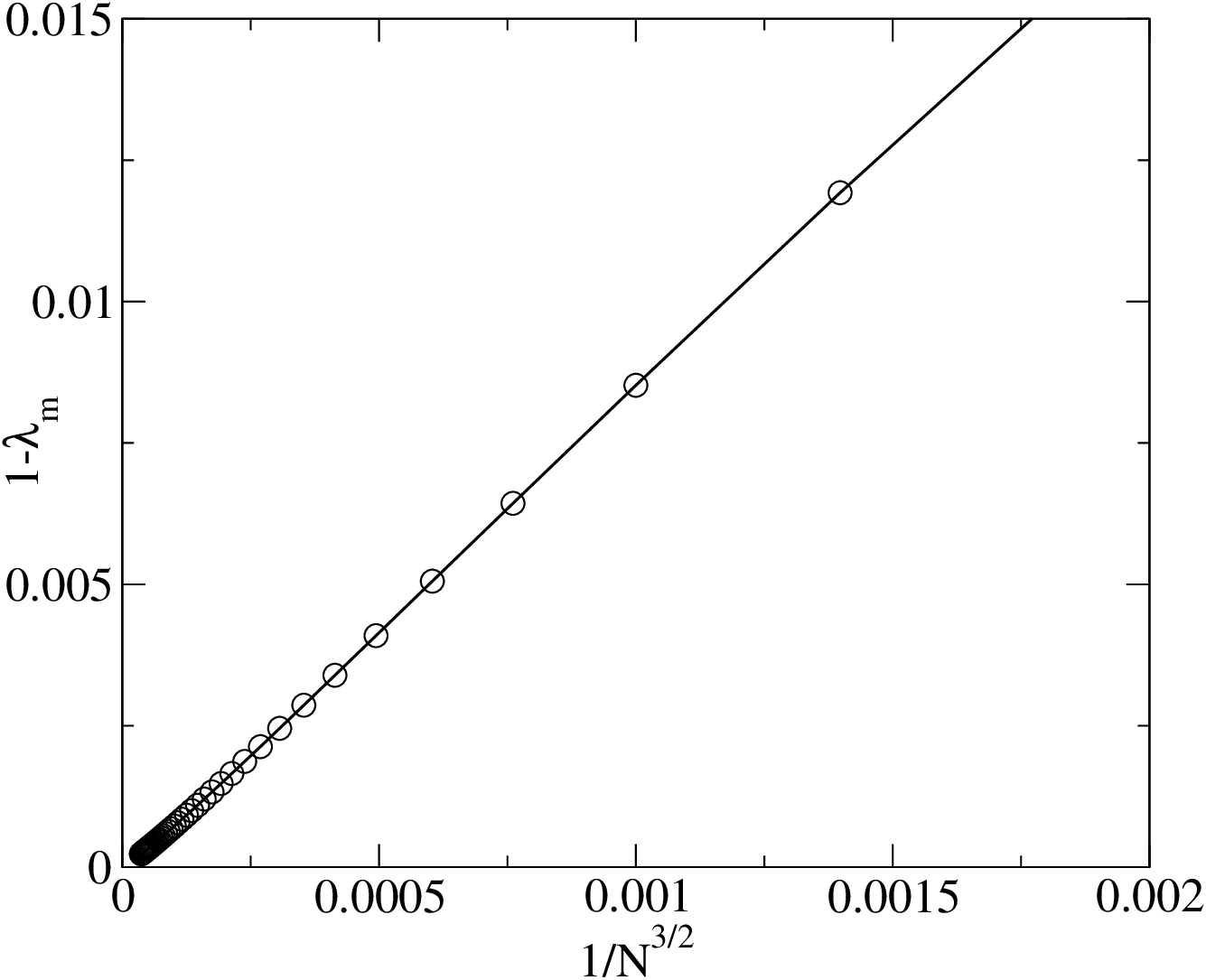}
\caption{\label{figure_lambdam} The scaling behavior of $\lambda_m$.}
\end{figure}

\begin{figure}
\includegraphics[width=8cm]{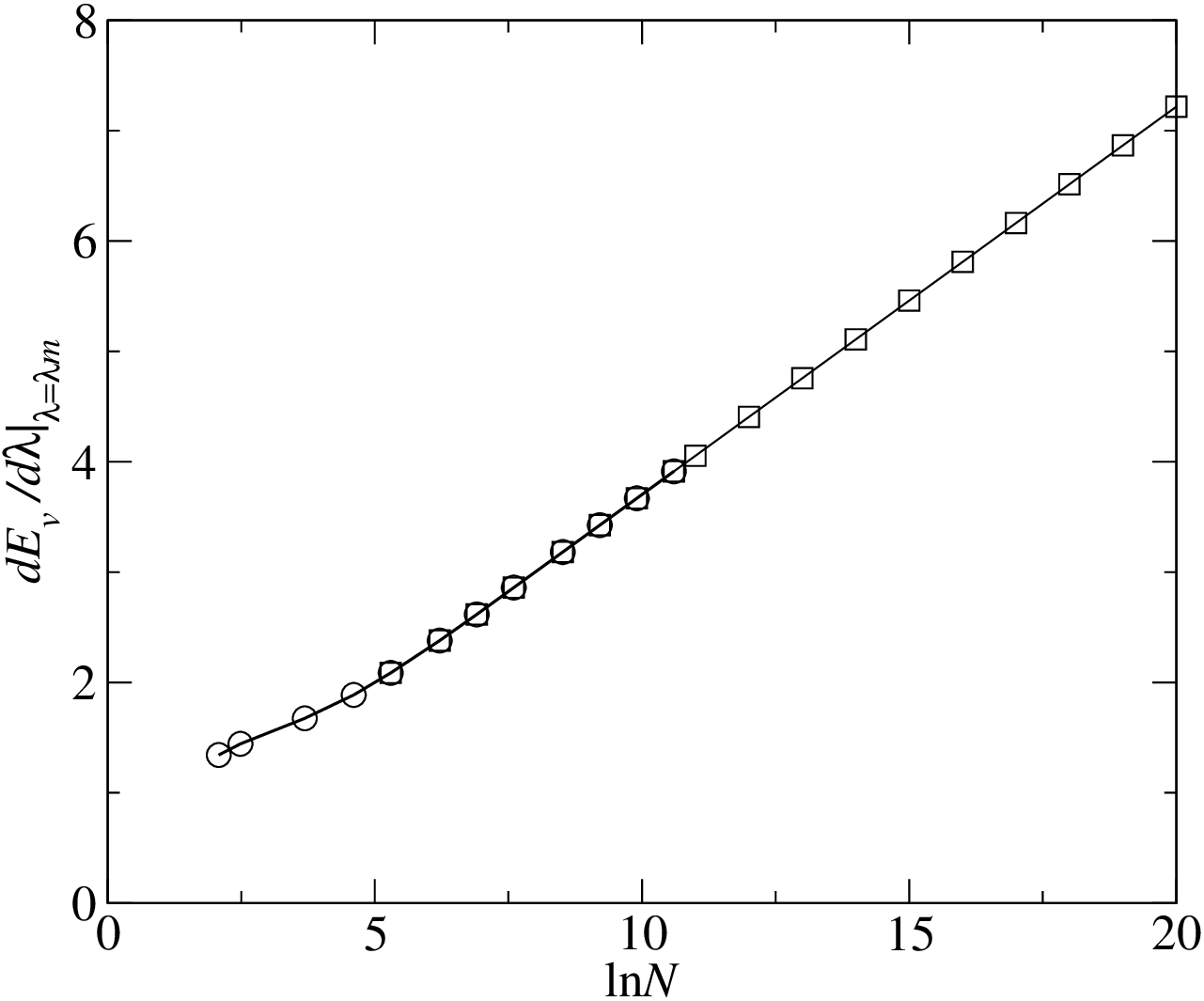}
\caption{\label{figure_dlentm} The scaling behavior of the maximum point of
$dE_v/d\lambda$. The circle line is obtained by numerical derivative, and the
square line by analytical expression directly.}
\end{figure}

As is well known, in a ring with infinite sites, the phase diagram of the
transverse field Ising model is divided into two phases by a transition point
$\lambda=1$. For $\lambda<1$, the ground state of the system is a paramagnet.
Its physical picture can be understood from the case of $\lambda=0$ at which all
spins are fully polarized along $z$ direction. While for $\lambda > 1$, the
strong Ising coupling introduces magnetic long-range order of the order
parameter $\langle\sigma^x\rangle$ to the ground state. Therefore, the
transition in this model is of type from order-to-disorder, and is quite
different from the transition in some other models. In the 1D XXZ model, it has
been shown that the two-site local entanglement reaches a local maximum at the
transition point $\Delta=1$ which witnesses a transition of type from
order-to-order, and becomes singular at another transition point $\Delta=-1$ due
to the ground-state level-crossing. However, in the 1D Ising model, no similar
properties can be observed directly from the entanglement, as shown in Fig.
\ref{figure_lent}. As suggested by the classification scheme for the quantum
phase transition based on the pairwise entanglement \cite{SJGu0511243}, we take
the first order derivative of the entanglement with respect to the coupling
$\lambda$. The results are shown in Fig. \ref{figure_dlent}. We are happy to
notice that the first order derivative around the critical point becomes sharper
as the system size increases, and is expected to be divergent in an infinite
system.

In the critical phenomena, the most important themes are the scaling and
universality. In a finite sample, a physical quantity is a smoothly continuous
function if there is no ground-state level-crossing. However, it is expected
that the anomalies will become clearer and clearer as the size of sample
increases. The relevant study is the finite size scaling. For this
system, in order to study the scaling behavior the maximum point of
$dE_v/d\lambda$, we locate the maximum point for a given system size numerically
and define the corresponding location as $\lambda_m$. From Fig.
\ref{figure_dlent}, though there is no divergence when $N$ is finite, the
anomalies are obvious. The position of the maximum point $\lambda_m$ scales as
$\lambda_c-\lambda_m\propto 1/N^{3/2}$.

The results of $dE_v/d\lambda$ at $\lambda_m$ up to 2000 sites are shown in Fig.
\ref{figure_dlentm}. Though we can almost conclude logarithmic divergence of
$dE_v/d\lambda$ at the transition point from the figure, it is still very
difficult to take numerical derivative for a very large system, such as
$N=10^8$. In order to make it more confirmative, we now give a rigorous prove
that $dE_v/d\lambda$ obeys a logarithmic behavior at the critical point.

\begin{figure}
\includegraphics[width=8cm]{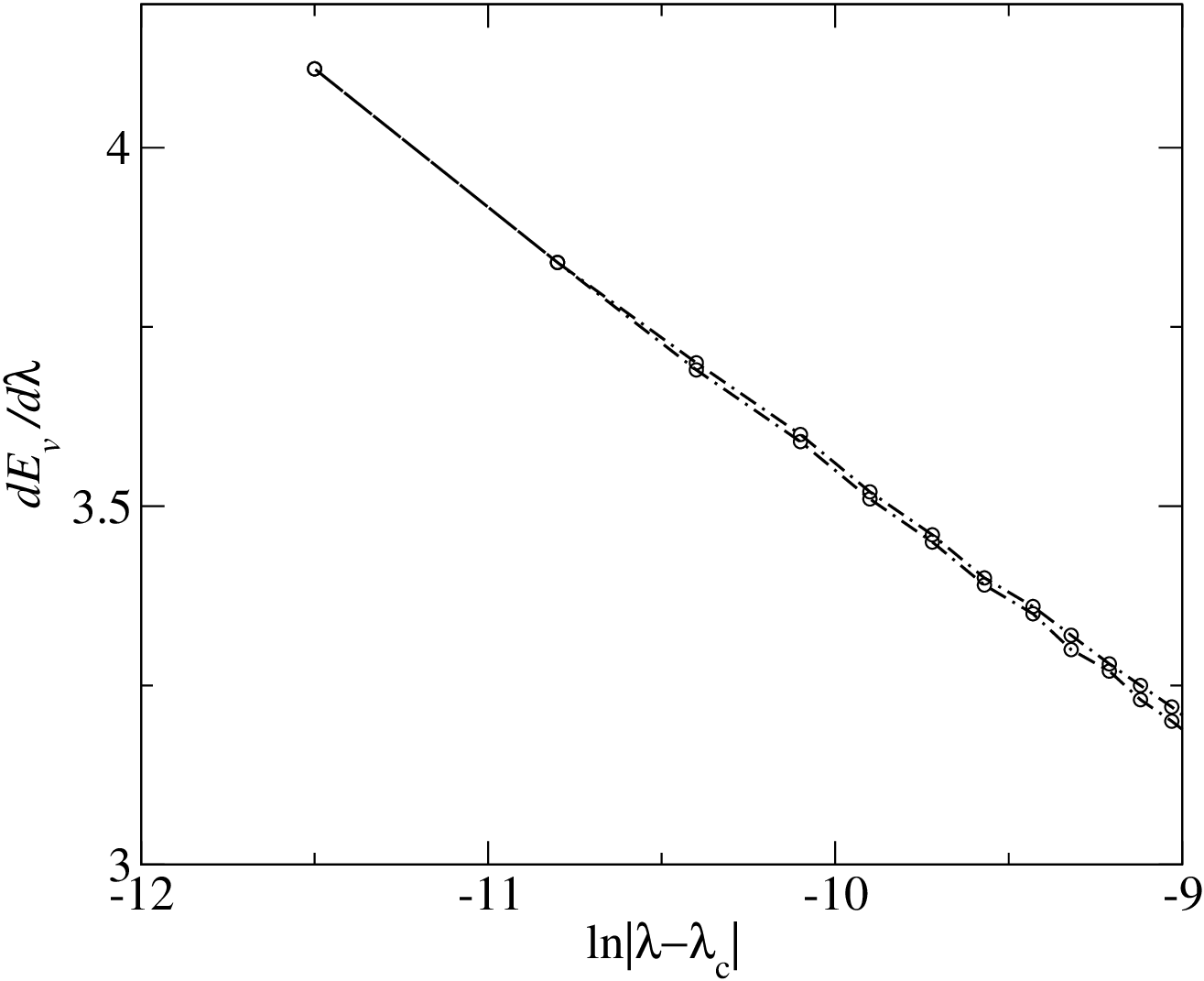}
\caption{Logarithmical divergence of $dE_v/d\lambda$ around the critical point
of an infinite system.\label{figure_logd}}
\end{figure}

In order to calculate the entanglement, we first need to calculate the
eigenvalues of the reduced density matrix
\begin{widetext}
\begin{eqnarray}
\epsilon_{1,2}&=&\frac{1}{4}\left[(1+\langle\sigma^z\rangle^2
-\langle\sigma^x_0\sigma^x_1\rangle\langle\sigma^y_0\sigma^y_1\rangle)
\pm\sqrt{4\langle\sigma^z\rangle^2+(\langle\sigma^x_0\sigma^x_1\rangle
-\langle\sigma^y_0\sigma^y_1\rangle)^2}\right], \nonumber \\
\epsilon_{3,4}&=&\frac{1}{4}\left[(1-\langle\sigma^z\rangle^2
+\langle\sigma^x_0\sigma^x_1\rangle\langle\sigma^y_0\sigma^y_1\rangle)
\pm(\langle\sigma^x_0\sigma^x_1\rangle+\langle\sigma^y_0\sigma^y_1\rangle)\right]
\end{eqnarray}
\end{widetext}
Then the first order derivative of the entanglement takes
\begin{eqnarray}
\frac{dE_v}{d\lambda}=-\log_2\frac{\epsilon_{4}}{\epsilon_{3}}
\frac{d\epsilon_{4}}{d\lambda}-\log_2\frac{\epsilon_{3}}
{\epsilon_{1}}\frac{d(\epsilon_{3}+\epsilon_{4})}{d\lambda}
-\log_2\frac{\epsilon_{2}}{\epsilon_{1}}\frac{d\epsilon_{2}}{d\lambda}.
\end{eqnarray}
At the critical point, the eigenvalues $\epsilon_i$ converges quickly as the
system size increases
\begin{eqnarray}
\epsilon_{1,2}=\frac{1}{4}+\frac{4}{3\pi^2}\pm\frac{\sqrt{13}}{3\pi},\;\;\;
\epsilon_{3,4}=\frac{1}{4}-\frac{4}{3\pi^2}\pm\frac{1}{3\pi}.
\end{eqnarray}
Then after some complicated calculations, one can obtain
\begin{eqnarray}
\frac{dE_v}{d\lambda}&=&-\frac{1}{2}\left(\frac{1}{N}\sum_{\phi}
\left|\frac{\cos^2\frac{\phi}{2}}{\sin\frac{\phi}{2}}\right|
\cos\phi\right)\log_2\frac{\epsilon_{4}}{\epsilon_{3}}\\
&+&\left[-\left(\frac{1}{N}\sum_{\phi}\left|\frac{\cos^2\frac{\phi}{2}}
{\sin\frac{\phi}{2}}\right|\right)\left(\frac{1}{N}\sum_{\phi}\left|
2\sin^2\frac{\phi}{2}\right|\right)\right.\nonumber
\\
&+&\left.\frac{1}{N^2}\left(\sum_{\phi}\left|4\cos^2\frac{\phi}{2}\sin\frac{\phi}{2}
\right|\right)^2\right]\frac{1}{2(\epsilon_{1}-\epsilon_{2})}\log_2\frac{\epsilon_{2}}{\epsilon_{1}}.
\nonumber
\end{eqnarray}
The main contribution to the above expression in the large $N$ limit arise from
the summation around zero point of $\phi$. Because
\begin{eqnarray}
\frac{1}{N}\sum_{\phi \in S}\frac{1}{|\frac{\phi}{2}|}\simeq
-\frac{2}{\pi}\ln\frac{2\pi}{N}\simeq \frac{2}{\pi}\ln N ,
\end{eqnarray}
then we obtain, in the large $N$ limit
\begin{eqnarray}
\frac{dE_v}{d\lambda}\simeq A_{1}^E \ln N
\end{eqnarray}
where $A_{1}^E$ is a constant
\begin{eqnarray}
A_{1}^E &=& -\frac{1}{2\pi}\log_2{\frac{3\pi^2+4\pi-16}{3\pi^2-4\pi-16}} \nonumber \\
&& +\frac{3}{2\sqrt{13}\pi}\log_2{ \frac{3\pi^2+4\sqrt{13}\pi+16}
{3\pi^2-4\sqrt{13}\pi+16} }
\end{eqnarray}
as has been shown in Fig. \ref{figure_dlentm}. On the other hand, for an
infinite system, $dE_v/d\lambda$ diverges on approaching the critical point as
\begin{eqnarray}
\frac{dE_v}{d\lambda}=A^E_2\ln|\lambda-\lambda_c|+{\rm const},
\end{eqnarray}
as has been shown in Fig. \ref{figure_logd}. Here $A^E_2$ is not completely
independent of $\lambda$, and $A^E_2= A^E_1$ at $\lambda=1$.

\begin{figure}
\includegraphics[width=8cm]{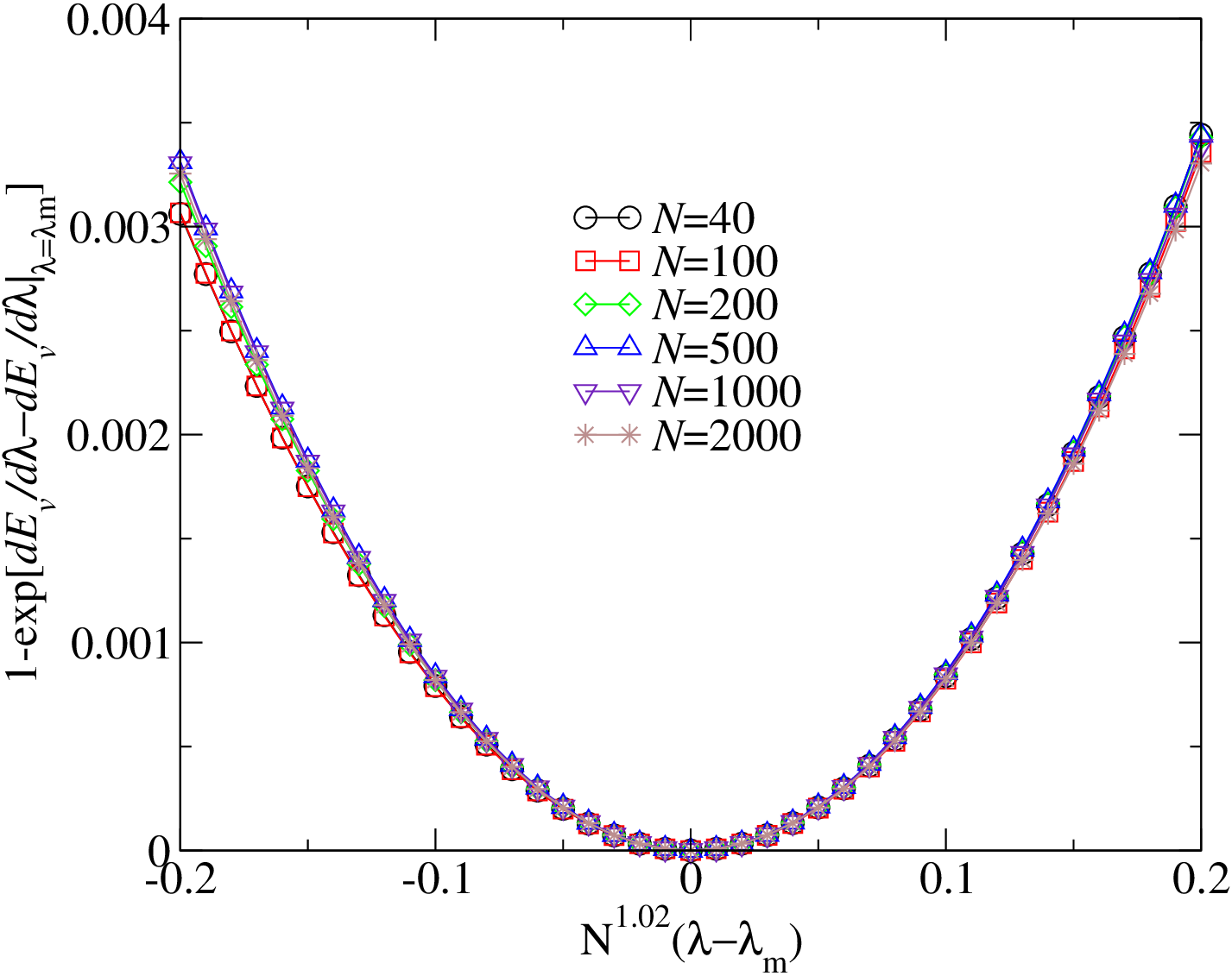}
\caption{\label{figure_scalf} (color online) The finite-size scaling analysis
for the case of logarithmic divergence. The local entanglement, considered as a
function of system size and the coupling, collapse on a single curve for
various system size.}
\end{figure}

According to the scaling ansatz \cite{MNBarberb}, the two-site entanglement,
considered as a function of the system size and the coupling, is a function of
$N^{1/\nu}(\lambda-\lambda_m)$. In the case of logarithmic divergence, it
behaves as $dE_v/d\lambda - dE_v/d\lambda|_{\lambda=\lambda_m} \sim
Q[N^{1/\nu}(\lambda-\lambda_m)]$, where $Q(x)\propto \ln x$ for large $x$. In
Fig. \ref{figure_scalf}, we perform the scaling analysis and find the
entanglement can be approximately collapse to a single curve for $\nu\simeq
0.98$. This behavior is quite different from that of
concurrence,\cite{AOsterloh2002} which diverges as
$dC/d\lambda|_{\lambda=\lambda_m}\propto-0.2702\ln N$ and
$dC/d\lambda|_{\lambda=\lambda_c}\propto (8/3\pi^2)\ln |\lambda-\lambda_c|$
respectively. Because of $8/3\pi^2\simeq 0.2702$, the ratio of constants in the
ln term is unit. This fact leads to $\nu=1$ for the
concurrence.\cite{AOsterloh2002} Actually, though Osterloh {\it et al} only
gave a numerical constant 0.2702 for the logarithmical divergence of the
concurrence with increasing system size, it also can be exactly obtained that
the constant is just $8/3\pi^2$ with the above procedure. However, for the
local entanglement, since $A^E_2$ slightly depends on $\lambda$, we have
$\nu\simeq 0.98$ according to numerical analysis.

In summary, we have investigated the scaling behavior of the two-site local
entanglement in a quantum phase transition of 1D transverse field Ising
model. Different from the entanglement in the transition of 1D XXZ model at
$\Delta=1$ which is of type from order-to-order, the entanglement in 1D
transverse field Ising model is not a maximum at the critical point where the
quantum phase transition is of the type from order-to-disorder. However, its
first order derivative with respect to the coupling becomes singular around the
critical point as the system size increases.  We show the logarithmic
divergence both numerically and analytically.

{\it Note added:} During the preparation of the manuscript, we notice that a
work on the two-site entanglement in similar model on the arXiv \cite{HDChen06}.

This work is supported by the Earmarked Grant for Research from the Research
Grants Council of HKSAR, China (Project CUHK N\_CUHK204/05 and HKU\_3/05C) and
Direct Grant of CUHK (A/C 2060286).


\begin{references}
\bibitem{ABinstein35}
A. Einstein, B. Podolsky, and N. Rosen, Phys. Rev. {\bf 47}, 777 (1935).

\bibitem{SeeForExample}
See, for example, C. H. Bennett, D. P. Divincenzo, Nature {\bf 404}, 247 (2000).

\bibitem{MANielsenb}
M. A. Nielsen, and I. L. Chuang, {\it Quantum Computation and Quantum
Information} (Cambridge University Press, Cambridge, 2000).



\bibitem{SJGUPRL}
S. J. Gu, S. S. Deng, Y. Q. Li, H. Q. Lin, Phys. Rev. Lett. {\bf 93}, 086402
(2004); S. S. Deng, S. J. Gu, and H. Q. Lin, Phys. Rev. B {\bf 74}, 045103
(2006).


% two-point versus multipartite entanglement in quantum phase transitions
\bibitem{anfossi2005}
A. Anfossi, P. Giorda, A. Montorsi, and F. Traversa, Phys. Rev. Lett., {\bf 95},
056402 (2005).

% Entanglement scaling in the 1d Hubbard model at criticality
\bibitem{larsson2005}
D. Larsson, and H. Johannesson, Phys. Rev. Lett. {\bf 95}, 196406 (2005).

% single-site entanglement at the superconductor-insulator transition in the Hirsch model
\bibitem{anfossi2006}
A. Anfossi, C. D. E. Boschi, A. Montorsi, and F. Ortolani, Phys. Rev. B {\bf
73}, 085113 (2006).

% Two-site entropy and Quantum Phase Transitions in Low-Dimensional Models
\bibitem{legeza2006}
\"{O}. Legeza, and J. S\'{O}lyom, Phys. Rev. Lett. {\bf 96}, 116401 (2006).

\bibitem{gu2006le}
S. J. Gu, G. S. Tian, and H. Q. Lin, New J. of Phys. {\bf 8}, 61 (2006).

%local measures of entanglement and critical exponents at quantum phase transitions
\bibitem{venuti2006}
L. C. Venuti, C. D. E. Boschi, M. Roncaglia, and A. Scaramucci, Phys. Rev. A
{\bf 73}, 010303(R) (2006).

%two reference on the block-block entanglement
\bibitem{GVidal2003}
G. Vidal, J. I. Latorre, E. Rico, and A. Kitaev, Phys. Rev. Lett. {\bf 90},
227902 (2003).

\bibitem{VEKorepin04}
V. E. Korepin Phys. Rev. Lett. {\bf 92}, 096402 (2004).


\bibitem{JCao06}
J. Cao, W. Lu, Q. Niu, Y. Wang, cond-mat/0605460.

%Quantum Phase Transition
\bibitem{Sachdev}
Sachdev S. 2000 {\it Quantum Phase Transitions} (Cambridge University Press,
Cambridge, UK)

\bibitem{YChenNJP}
Y. Chen, P. Zanardi, Z. D. Wang, and F. C. Zhang, New J. Phys. {\bf 8} 97 (2006).

%Mean magnetization in the transverse field Ising model
\bibitem{EBarouch70}
E. Barouch, and B. M. McCoy, Phys. Rev. A {\bf 2}, 1075 (1970).

%Correlation function in the transverse field Ising model
\bibitem{EBarouch71}
E. Barouch, and B. M. McCoy, Phys. Rev. A {\bf 3}, 786 (1971).

%Entanglement and quantum phase transition
\bibitem{SJGu0511243}
S. J. Gu, G. S. Tian, and H. Q. Lin, quant-ph/0511243.


%Scaling ansatz on the logarithmic divergence.
\bibitem{MNBarberb}
M. N. Barber, {\it Phase Transitions and Critical Phenomena} {\bf 8}, 146-259
(Academic, London, 1983).

\bibitem{AOsterloh2002}
A. Osterloh, Luigi Amico, G. Falci and Rosario Fazio, Nature {\bf 416}, 608
(2002).


\bibitem{HDChen06}
Han-Dong Chen, cond-mat/0606126.

\end{references}
\end{document}